\documentclass[preprint,12pt]{elsarticle}
\usepackage{amssymb}
\usepackage{amsmath}
\usepackage{subfigure}
\journal{Wave Motion}
\usepackage{verbatim}

\begin{document}

\begin{frontmatter}

\title{Guided Wave Propagation in Complex Curved Waveguides I: Method Introduction and Verification}

\author{E. Khajeh, L. Breon, J. L. Rose}

\address{Engineering Science \& Mechanics Department, The Pennsylvania State University, University Park, PA 16802}
\begin{abstract}
A phenomenological method is developed to consider elastic guided wave propagation in complex curved waveguides. The theory on guided wave propagation in hollow circular cylinders is used in order to verify the method. The results are compared with solutions obtained by the Helmholtz decomposition method. The method given here is used to derive new analytical relations for torsional and longitudinal phase velocity dispersion curves. In addition, a new excitation method is proposed and examined for flexural modes. Some conceptual results and applications of the method are discussed.
\end{abstract}

\begin{keyword}
Guided wave \sep Ray-plate method\sep Pipe\sep Elbow \sep Flexural modes \sep Dispersion curves \sep Guided wave excitation \sep Curved plate \sep Curved waveguide
\end{keyword}

\end{frontmatter}
\section{Introduction}
\label{Intro.}
Ultrasonic guided waves are utilized for rapid, long-range inspections of various structures even when access to the structure is limited \cite{Rose-2002-baseline}. Guided waves are used extensively for inspecting isotropic single layer plates \cite{Chimenti-1997}, composite plates \cite{Su-2006}, pipelines \cite{Alleyne-1998,Shin-1999}, rail roads \cite{Rose-2002}, adhesive joints \cite{Heller-2000}, and more. In addition, current industrial demand for applications of guided waves is growing. Often, new applications of guided waves involve complex structures.\\
Guided waves are more complicated than bulk waves because of their dispersive nature, infinite number of propagating modes, wave velocity and wave structure variations with mode and frequency, and waveguide boundary conditions \cite{rose,Achenbach,Auld}. Despite the complex nature of guided waves, a complete understanding of guided wave propagation in a waveguide is necessary for robust theoretically driven inspection. For example, a prior knowledge of group velocity,  wave structure, and skew angle of excited modes is required to estimate the location and type of a defect in a composite plate. On the other hand, if the complexities of guided waves are quantitatively understood and utilized, the sensitivity and penetration power of guided waves can be significantly improved for inspection purposes . For instance, wave propagation characteristics of non-axisymmetric flexural modes in a pipe are used for synthetic and active focusing \cite{Hayashi-2005,Rose-2003}. Another example of how theoritically understanding of guided waves can be applied to improve inspections is using wave structure matching for inspection of adhesive joints \cite{Lowe-1994}.\\
Guided wave dispersion curves are crucial for presenting the phase and group velocity variation with frequency. Furthermore, dispersion curves reveal the dispersivity characteristics of an excited mode. This information is useful for signal post-processing. Dispersion curves are also used to determine suitable modes for excitation, the angles of angled beams, and the spacing of comb type transducers to excite special modes \cite{rose}.\\
While dispersion curves, wave structure, mode selection, scattering properties, and wave propagation characteristics of guided waves should be studied for efficient and successful inspection, these concepts can be considered in a straight forward fashion for some simple geometries like single layer plates, multilayer plates, and pipes,where Navier's equation can be solved by Helmholtz decomposition or by the partial wave method \cite{rose, Auld}. Finite element analysis can be used when considering complex waveguides, but the method suffers from a fundamental inability to provide a generic solution for further sophisticated implementation of guided waves. In addition, computational cost and time consumption of finite element analysis is high, so it cannot be used for geometrically large waveguides. Semi-analytical finite element analysis (SAFE) can decrease the computation cost but  is applicable for structures only with constant, albeit arbitrary cross section \cite{Hayashi}. \\
On the other hand, understanding wave propagation properties in some important but complex curved geometries is highly desirable. For example, elbows are an essential element of pipelines, so an effective long range inspection of a pipeline depends strongly on the ability to inspect beyond elbows \cite{Hayashi-2005-2}. The understanding of wave propagation in an elbow should allow data interpretation of reflected waves from beyond the elbow. A growing number of NDE projects are becoming of interest in which complex curved surfaces are involved. Analytical methods like Helmholtz decomposition and the partial wave method cannot provide solutions for geometrically complex curved structures; and applicability of numerical solutions is also limited. Therefore, developing a practical method to consider guided wave propagation in complex curved waveguides would be highly beneficial for extending the applications of guided waves for inspection purposes.\\
In this work, a physical understanding of guided waves propagation is presented. A method for predicting guided wave propagation in complex waveguides is developed by considering the current understanding of different aspects of wave propagation. Then, upon the new understanding given here, a quantitative method is developed.  The method is used to derive an analytical relation for  torsional and longitudinal phase velocity dispersion curves in a hollow circular cylinder. The results of the method are compared with other available solutions. Limitations of the method are considered in section \ref{Limitations}. In section \ref{FME}, a new method for flexural mode excitation is proposed based on a newly developed perspective of guided wave propagation. Finite Element models are used to show that the method effectively excites the desired flexural modes. A summary of the method and potential applications are considered later in the text.
\section{Method Description}
\label{sec:M}
\begin{figure}[h]
\centering
\includegraphics[height=.5\textheight,width=1\textwidth]{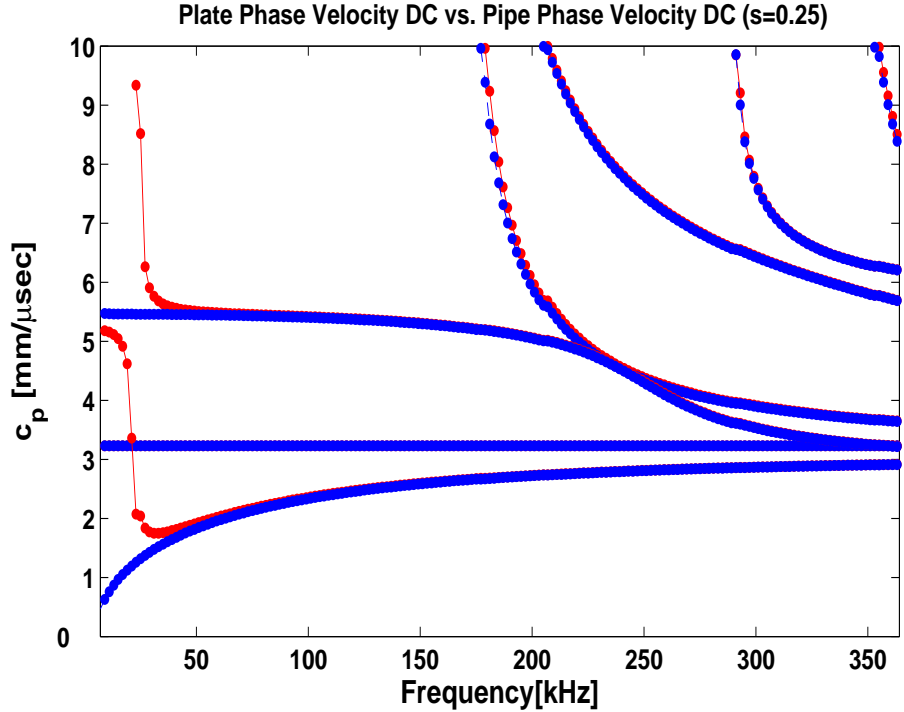}
\caption{Figure shows the axisymmetric dispersion curves in a steel pipe with $ s=0.25 $ (red) and the dispersion curves in a steel plate (blue). The diagram shows that the axisymmetric dispersion curves in the pipe are nearly the same as the dispersion curves in the plate for higher frequencies.}
\label{1}
\end{figure}

A method is developed to study wave propagation in complex curved waveguides. There are a few approximation methods that can simplify the wave equation by low or high frequency approximations. They are based upon a Taylor series expansion of the wave equation and then eliminating the higher orders based on some assumptions \cite{Harris-2000,Babich-2005,Tanner-2007,Gregory-1970}. In this work, a phenomenological approach is adopted instead of a mathematical approach. It is shown that the method is accurate and simple to apply in practical situations. The method is motivated by the following facts.
\begin{enumerate}
\item A plane wave in an infinite medium is described by a specific direction of propagation and also a sinusoidal change in amplitude which is a function of time and distance along the wave propagation direction. One wavevector is chosen on behalf of an infinite number of wavevectors that exists all around the propagation region, because all of the wavevectors have the same direction and amplitude in the case of plane wave. Consider the generalized case where the wave propagation direction is different at each point. Then the plane wave symmetry is broken and the wave should be determined by a variable vector field that is a natural generalization of constant wavevectors.
\item The plane wave in infinite medium moves on a straight line, because the medium is assumed to be isotropic and flat. Consider a plane wave that moves in a two dimensional flat plate. The propagation path of the plane wave is considered as a straight line, but if the plate is curved the propagation path changes to a curved line.
\end{enumerate}
Therefore, in order to extend a plane wave propagation in an isotropic medium to a multi-directional wave propagation on a two dimensional curved surface, the above facts lead us to the following perspective.
\begin{enumerate}
\item An arbitrary wave is defined by a vector field that is changing point by point according to the underlying geometry and initial emission conditions.
\item An infinite number of paths can be defined on the surface that are tangential to the propagation direction vector field at each point. These paths are called \emph{ray-paths}. In fact, ray-paths on a surface are geodesics of the surface. Geodesics are defined as straight lines on a curved surface. The geodesics of a surface can be determined by the metric of the surface and the following relation \cite{spivak}:
\begin{equation}\label{g1}
\frac{dx^\lambda}{ds^2}+\Gamma^\lambda_{\mu\nu}\frac{dx^\mu}{ds}\frac{dx^\nu}{ds}=0,
\end{equation}
where $ s $ is affine parameter, $\lambda,\mu,\nu=1,2$ and $\Gamma^\lambda_{\mu\nu}$'s are the Christoffel symbols of the metric.\\
\item The intensity of the wave will change sinusoidally along each ray-path.
\end{enumerate}
We call the method a \emph{ray-method}. The method can be summarized as follows. Assume that the excitation from a certain source on a complex curved surface is studied. Infinite emitted rays are considered where the excitation conditions determines the initial position, initial direction, and maximum intensity of each ray. Then, the geodesic equation determines the propagation path of each ray on the surface. The intensity of each ray changes sinusoidally along the ray-path. In order to find the intensity of the wave at each point, a superposition of all ray-path intensities that pass through the point at the same time is considered.\\
In order to develop a complete method of studying guided wave propagation in a curved waveguide, though, the effect of thickness on guided waves should be addressed. Until now, we have used rays to consider guided wave propagation. In particular, rays that lie on the surface of the waveguide. In this model, the thickness of the waveguide cannot influence ray propagation. Indeed, the ray-method cannot capture guided waves properties caused be the thickness of the waveguide like dispersion curves and wave structures. Now, the dispersion curves of a plate are compared with the dispersion curves of a pipe in order to get some insight on how the thickness can be included in the ray-method.\\
Two phase velocity dispersion diagrams are defined for a plate:  shear-horizontal dispersion curves (SHDCs) and Rayleigh-Lamb dispersion curves (RLDCs). SHDCs are derived from the following analytical relation \cite{rose}
\begin{equation}\label{eq:SH}
c_p^{(n)}(fd)=\frac{2c_T(fd)}{\sqrt{4(fd)^2-n^2c_T^2}}.
\end{equation}
Where $c_T$, $fd$, and $c_p^{(n)}$ are the shear wave velocity, the frequency-thickness product, and the phase velocity of the nth mode, respectively. RLDCs are derived numerically from the Rayleigh-Lamb transcendental equation \cite{rose}. Governing transcendental equations for deriving dispersion curves in a pipe are more complicated than for a plate \cite{gazis}. The complicated transcendental equations need to be solved in order to find torsional  dispersion curves (TDCs) and longitudinal dispersion curves (LDCs) in a pipe. Figure \ref{1} shows the axisymmetric part of TDCs and LDCs in a steel pipe (red) with $ s=0.25 $ \footnote[1]{$ s $ is the ratio of pipe thickness to pipe mean radius} and  SHDCs and RLDCs in a steel plate (blue) on the same graph. Considering the dispersion curves in Figure $  $\ref{1} reveals that the axisymmetric part of pipe dispersion curves nearly are the same as plate dispersion curves for higher frequencies. The primary differences are in the low frequency limit where $ L(0,1) $ and $ L(0,2) $ experience a jump to higher phase velocities.  This jump for the pipe with $ s=0.25 $ occurs around $ 25[kHz] $. The frequency where this jump occurs tends toward zero as the $ s $ factor decreases. In most practical pipes the $ s $ parameter is less than $ 0.25 $, so we can approximate the axisymmetric part of pipe dispersion curves by plate dispersion curves.\\
An axisymmetric mode in the context of the ray-method is represented by rays that move along the pipe axial direction. Thus, the similarity between the axisymmetric part of the pipe dispersion curves and the plate dispersion curves means that we should get plate type dispersion in a pipe, if we consider a set of rays  moving along the pipe axial direction. This observation leads to the notion that instead of using a single ray confined to the surface of a waveguide we could consider a curved plate spanning the thickness of the waveguide where the plate falls along a ray-path and is normal to both free boundaries of the waveguide. In this context, the axisymmetric dispersion curves of a pipe will be same as  the dispersion curves of a plate. Hereafter, this concept will be referred to as \emph{ray-plate} method.\\
In the following section, guided wave propagation in a pipe is studied by the ray-plate method. New analytical relations are derived for pipe dispersion curves using the ray-plate method. Results from the ray-plate method are compared with exact results derived by the global matrix method. It is shown that the ray-plate method predicts accurate results for the pipe dispersion curves. In addition, it provides new insight on pipe flexural modes.
\section{Verification of Ray-Plate Method}
\label{MV}
\begin{figure}
\centering
\mbox{
\subfigure[]{
   \includegraphics[height=.4\textheight,width=.9\textwidth] {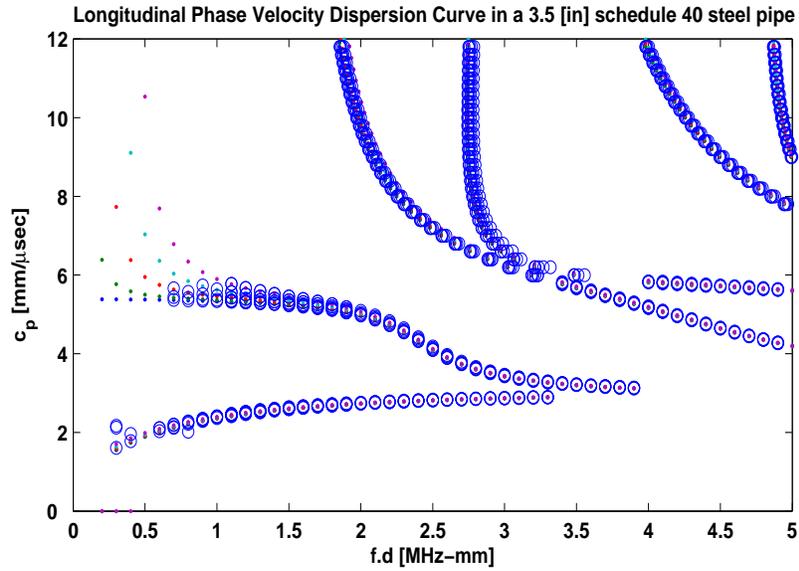}
   \label{fig:LDCsubfig1}
 }
 }
 \mbox{
  \subfigure[]{
	\includegraphics[height=.4\textheight,width=.9\textwidth] {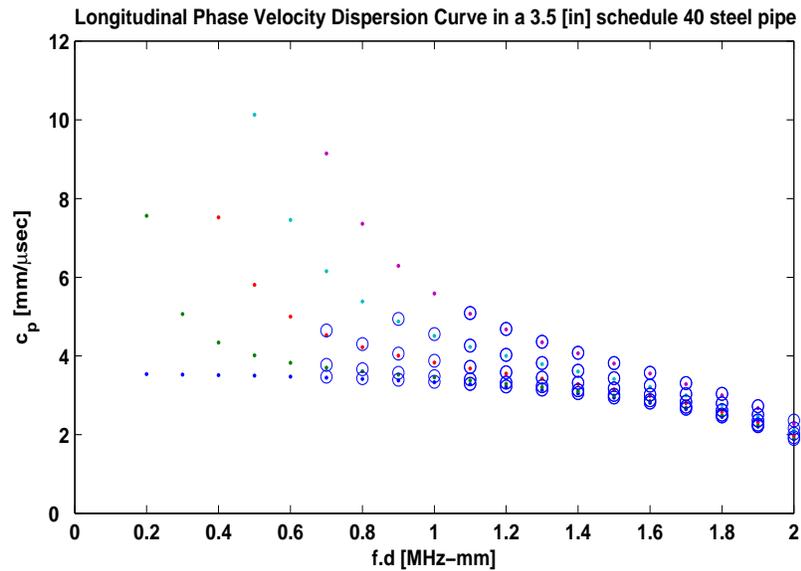}
	\label{fig:LDCsubfig2}
  }
}
\label{figure:LDC1}
\caption{LDCs derived by ray-plate method and by global matrix are compared in these diagrams. a) compares the result of the ray-plate method and the global matrix method for 3.5[in] schedule 40 steel pipe , b) shows a magnified view of fig.(a) for lower frequencies. The figures present an excellent agreement between results of the ray-plate method and the global matrix method.   }
\end{figure}

\begin{figure}[ht]
\centering
\includegraphics[height=.35\textheight,width=.9\textwidth] {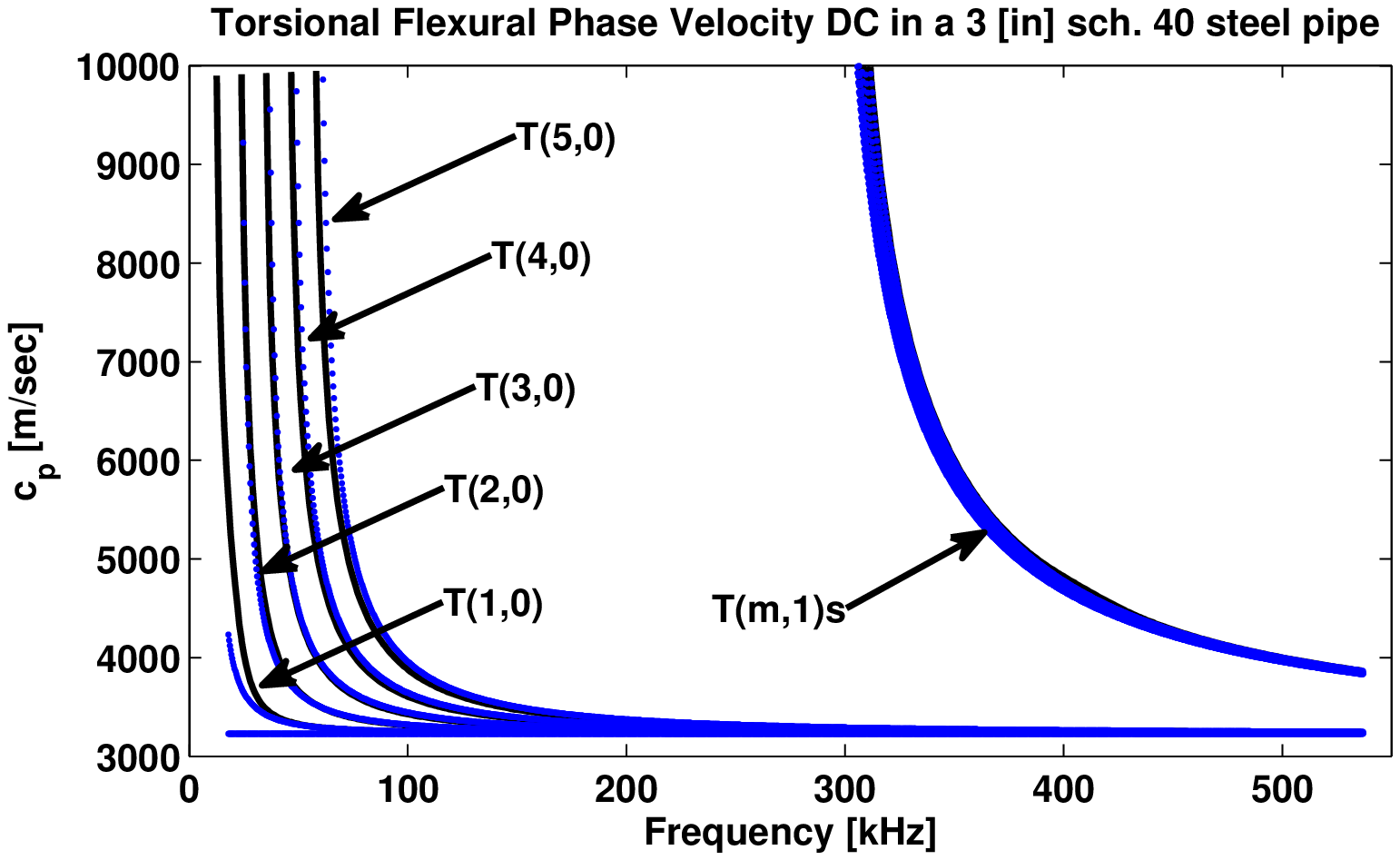}
\label{fig:TDC}
\caption{ This figure presents a comparison between the results of the ray-plate method and the global matrix method. It demonstrates that the ray-plate method closely approximates torsional dispersion curves for 3 [in] schedule 40 steel pipe.}
\end{figure}

In this section, the ray-plate method is applied to hollow circular cylinders. Guided waves in hollow circular cylinders have been studied extensively \cite{gazis}. The global matrix method with numerical calculation are often used in a complicated form to derive the pipe dispersion curves \cite{rose}. There are no analytical solutions for the torsional and the longitudinal dispersion curves of pipe. The ray-plate method is presented in detail and its results are compared with existing solutions for guided waves in hollow circular cylinders.\\
As mentioned before, comparing dispersion curves between plates and pipes show that SHDCs in plates are approximately the same as the axisymmetric TDCs in pipes and also that the RLDCs in plates approximate axisymmetric LDCs in pipes. Also, the axisymmetric modes in the context of the ray-method corresponds to an infinite set of rays that move parallel to the pipe axial direction. This leads to the extension of the ray-method from membranes to thick-walled shells, if instead of a set of rays, a set of plates across the thickness of the surface is considered. Each plate propagates along a ray-path and is perpendicular to both boundary surfaces of the shell. Each plate carries Lamb waves and/or SH waves. These plates are called \emph{ray-plates}.\\
The ray-plate method claims that axisymmetric modes are ray-plates that move parallel to the axial direction of pipe. Ray-plates that move at a relative angle with the axial direction can naturally become a candidate for explaining flexural modes. Let's construct dispersion curves in pipe by the ray-plate method base on the above insight about axisymmetric and flexural modes.\\
As mentioned above a flexural mode is constructed by a set of ray-plates carrying Lamb waves or SH waves and moving in a direction that makes an angle $\alpha$ with the pipe's axial direction (z-direction). Consequently, the phase velocity of the flexural modes and corresponding axisymmetric modes are the same at a fixed angular frequency $\omega$. The only difference is that these modes are moving in a different direction than axisymmetric modes, namely by the angle $\alpha$. Dispersion curves in a pipe are defined by choosing the z-axis as a preferred direction. The phase velocity of each mode is determined by $ c_p=\omega/k_z $, where $ k_z $ is the $ z $-component of the wavevector. Therefore, the phase velocity of a flexural mode can be derived by the following relation:
\begin{equation}\label{cp}
c_p^{(f)}=\frac{\omega}{k\cos(\alpha)}=\frac{1}{\cos(\alpha)}c_p^{(a)},
\end{equation}
where $c_p^{(a)}$ is the phase velocity of the axisymmetric mode at each frequency $\omega$ and $\alpha$ is the angle of wave propagation . The cylindrical shape of a pipe imposes a periodic displacement continuity boundary condition in the circumferential direction. A flexural mode is defined as a Lamb or SH wave existing in a ray-plate, where the ray-plate makes angle $\alpha$ with the z-axis. So, the displacement field components can be written as
\begin{equation}
\vec{u}(r,\phi,z)=\vec{U}(r)e^{i(k\cos(\alpha)z+k\sin(\alpha)R\phi-\omega t)}.
\end{equation}
In a pipe, if  $\phi$ is changed to $\phi+2\pi$, the displacement vector should be the same, meaning that $ \vec{u}(r,\phi,z)=\vec{u}(r,\phi+2\pi,z) $. This periodic condition imposes that
\begin{eqnarray}\label{PBC}
k\sin(\alpha)R=m \qquad   m=0,\pm 1,\pm 2,...,
\end{eqnarray}
where $ R $, $ k $, and $ m $ are the mean radius of the pipe, the wavenumber, and the flexural order. From \eqref{cp} and \eqref{PBC}, the following relation is established between the phase velocity of a flexural mode and the phase velocity of its corresponding axisymmetric mode at a given frequency,
\begin{equation}
c_p^{(f)}=\frac{1}{\sqrt{1-(\frac{m}{kR})^2}}c_p^{(a)}.
\end{equation}
The relation can be rewritten as
\begin{equation}\label{eq:cp1}
c_p^{(m,n)}(fd)=\frac{c_p^{(0,n)}}{\sqrt{1-\left[ \frac{msc_p^{(0,n)}}{2\pi (fd)}\right]^2}},
\end{equation}
where $c_p^{(m,n)}$ is phase velocity of a mode of family $n$ and flexural order $m$, $fd$ is the frequency-thickness product  and $s=d/R$ is the ratio of the pipe thickness to the pipe mean radius. Relation \eqref{eq:cp1} imposes the following condition on the maximum number of flexural modes that can exist at each frequency (f), Lamb mode phase velocity $c_p^{(0,n)}$ , and radius of the pipe (R):
\begin{equation}\label{eq:cp2}
m_{max}=int\left(\frac{\omega R}{c_p^{(0,n)}}\right)
\end{equation}
Eq.\eqref{eq:cp2} gives a new limitation on the number of flexural orders at each frequency. This relation can not be extracted simply from the global matrix method and needs further consideration to be recognized.\\
Eq.\eqref{eq:cp1} is used to derive LDCs in a pipe. Mean radius and thickness of pipe, plate phase velocities at each frequency, and flexural orders should be substituted in eq.\eqref{eq:cp1} in order to drive the LDCs.  Figure 2 shows a comparison between the LDC derived by the ray-plate method and the global matrix method. It is observed that at lower frequencies, flexural modes are separated more from each other than at higher frequencies. Figures \ref{fig:LDCsubfig1} and \ref{fig:LDCsubfig2} show that the derived LDCs in a 3.5[in] schedule 40 pipe by the ray-plate method are in good agreement with dispersion curves derived by the global matrix method.
The following analytical relation for TDCs, including flexural modes, can be derived by using equation \eqref{eq:cp1} and the shear horizontal dispersion relation \eqref{eq:SH}:
\begin{equation}\label{eq:TDC-modified}
c_p^{(m,n)}(fd)=\frac{2\pi C_T(fd)}{\sqrt{4 \pi^2(fd)^2-(\pi^2 n^2+ m^2 s^2)C_T^2}} ,
\end{equation}
where $ C_T $, $ fd $, $ n $, and $ m $ are shear wave velocity, frequency-thickness product, family order, and flexural order, respectively.  Figure \ref{fig:TDC} shows a comparison between dispersion curves of a 3[in] schedule 40 steel pipe derived by the ray-plate method and the global matrix method. The figure shows that the analytical relation \eqref{eq:TDC-modified} , derived by the ray-plate method, provides an accurate approximation for TDCs in pipes. \\
It is simple to check that if the limit $R \rightarrow \infty$ is considered, we get the shear horizontal dispersion curve relation. In addition, considering limit $(fd) \rightarrow \infty$ gives $c_p^{(m,n)}=C_T$. Therefore, equation \eqref{eq:TDC-modified} is a consistent equation in those limits. In addition, equation \eqref{eq:TDC-modified} imposes the following relation to existence of torsional mode $ (n,m) $ as
\begin{equation}\label{eq:TDC1}
n^2s^2+4m^2<\frac{4f^2R^2}{C_T^2}
\end{equation}
The ray-plate method gives some insight on why flexural mode group velocities are less than the axisymmetric mode group velocities. In the ray-plate method, flexural modes are related to rays whose propagation direction make an angle $\alpha$ with the z-axis. While the group velocities for all rays are the same, the propagation paths for the flexural modes will be longer than those of the axisymmetric modes. Therefore, the measured group velocities for flexural modes along the z-axis are less than those of their corresponding axisymmetric modes. Higher order flexural modes  correspond to longer paths and hence lower group velocities.

\section{Ray-Plate Method Limitations}

\label{Limitations}
\begin{figure}
\centering
\mbox{
\subfigure[]{
   \includegraphics[height=.35\textheight,width=.9\textwidth] {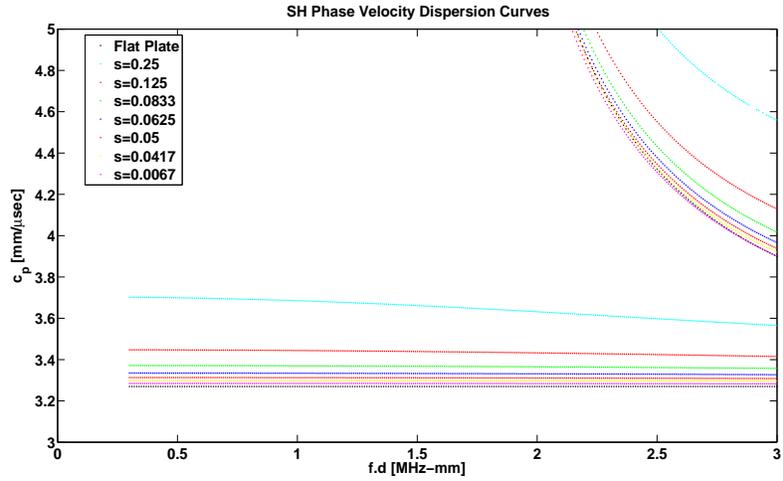}
   \label{fig:CPDCsubfig1}
 }
 }
\mbox{
 \subfigure[]{
   \includegraphics[height=.35\textheight,width=.9\textwidth] {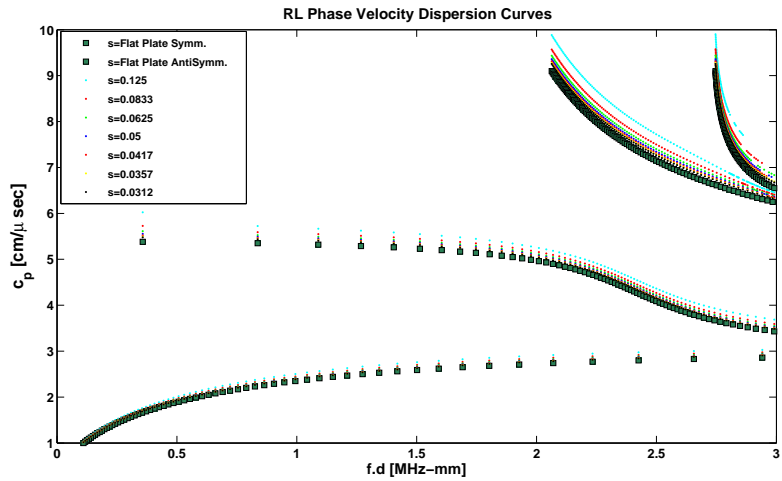}
   \label{fig:CPDCsubfig2}
 }
 }
\label{fig:CPDC}
\caption{Figures show dispersion curves for plates with different ratio of thickness to curvature radius ($ s $). a) shows SHDCs when $ s $ varies between $ 0.25 $ (cyan) and $ s_{plate}=0 $ (black).  b) shows RLDCs when $ s $ varies between $ 0.125 $ (cyan) and $ s_{plate}=0 $ ( green square). The diagrams show that for $ s<0.062 $, the dispersion curves of a curved plate can be approximated by the flat plate dispersion curves.}
\end{figure}

Limitations of the ray-plate method for the prediction of guided wave propagation in complex curved surfaces are now considered. It is shown that the ray-plate method presents a relatively accurate approximation when the thickness to curvature radius ratio of a structure is less than $ 0.062 $ at each point of the structure. Otherwise, some deviations from the exact solutions will be observed. In many practical situations, however, the  ratio of thickness to curvature radius is less than $ 0.062 $. Therefore, wave propagation in many practical curved surfaces can be considered by the ray-plate method. \\
Considering the derivation of pipe dispersion curves by the ray-plate method in the previous section showed that in the context of ray-plate method dispersion curves in a pipe can be derived from the dispersion curves of a plate. The ray-plates in general can be curved, for example, while axisymmetric modes in a pipe can only be constructed by flat ray-plates, the construction of flexural modes are possible only by using curved ray-plates. The curvature radius for the curved ray-plates in a pipe decreases for higher order flexural modes. The curvature radius for the axisymmetric ray-plate is infinite and for the flexural mode $ m=m_{max} $ is the radius of the cylinder.\\
For the sake of simplicity, the ray-plate method was based on flat ray-plates. Therefore, any deviation from a flat plate because of waveguide geometry can potentially affect the predictions of the ray-plate method. Here, phase velocity deviations of SHDCs and RLDCs in a curved plate from the flat plate dispersion curves are considered in detail. Then, we calculate an approximate turning point in which curvature of a curved ray-plate causes a significant deviation from the flat ray-plate and thereby invalidates the ray-plate method. A similar study has been done by Shao et al. \cite{Shao-2004} which is limited to the consideration of SH waves. Other aspects of the problem have also been studied by Viktorov \cite{Viktorov-1967} and Qu et al. \cite{Qu-1998} under the topic of guided circumferential waves in a circular annulus. They showed that the dispersion curves in a curved plate experience some deviations from flat plate dispersion curves; and that the deviations are higher for a smaller curvature radius.\\
A cylindrically curved surface (cylinder) is assumed to be long enough as to neglet edge effects. Guided waves are assumed to propagate in the circumferential direction. Navier's equation is solved in cylindrical coordinates. Then, SH and Rayleigh-Lamb dispersion curves are derived for a variety of thickness to curvature radius ratios.\\
Figures \ref{fig:CPDCsubfig1} and \ref{fig:CPDCsubfig2} show SH dispersion curves and Rayleigh-Lamb dispersion curves for plates with different thickness to curvature radius ratios. These figures reveal that for value $ s<0.062 $, the dispersion curves can be approximated by the plate dispersion curves. The $ s $ value for most of standard pipe sizes is less than $ 0.062 $. Therefore the ray-plate method can be applied accurately for most of them. The situation is the same for many other structures that do not have small curvature radii.

\section{Flexural Mode Excitation}
\label{FME}
\begin{figure}[ht]
\centering
\includegraphics[width=.47\textwidth] {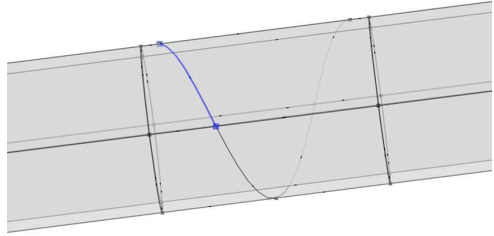}
\label{fig:loadsubfig2}
\caption{The new method of flexural mode excitation is presented in this figure. A helical edge load is applied on the pipe. The helix makes angle $ 20 [deg]$ with z-axis. So we expect to have a relatively pure excitation of flexural mode $ L(3,1) $.}
\end{figure}

Flexural modes in pipes  have been known about for decades \cite{gazis,Li-2001,Sun-2005}, but only basic physical understanding of these waves can be extracted from the direct solution of Navier's equation by the Helmholtz decomposition method. The ray-plate method provides a new physical understanding of these waves. The method reveals that flexural modes are basically like axisymmetric modes which propagate at an angle relative to the axial direction in a pipe. This new understanding of flexural modes suggests a new method of exciting pure flexural modes in pipe. In this section, the new flexural mode excitation method is proposed and examined by finite element analysis.\\
Flexural mode excitation has been considered by Li et al.  \cite{Li-2001} and Sun et al.  \cite{Sun-2005}, where partial normal loading and angle beam wedges are used to excite flexural modes. Those works used normal mode expansion to find the amplitude of the excited axisymmetric and flexural modes with respect to the characteristics of the load.  While flexural modes can be excited by these methods, they suffer two disadvantages. First, one pure flexural mode cannot be excited by these methods, several other flexural modes and also axisymmetric modes are excited at the same time. Next, the excited flexural modes can only partially cover the surface of a pipe, so a complete scan of the pipe is difficult  with these waves.\\
The active focusing technique is utilized for focusing a significant portion of guided waves energy at a desired point on the circumference position of a pipe. Li et al. \cite{Li-Jian-2005} and Sun et al. \cite{Sun-2005,Sun-2001,Sun-2002} used appropriate superposition of angular profiles of flexural modes at each axial distance in order to focus at the desired circumferential angle. Synthetic focusing is also used for defect imaging in pipe. Quasi-axisymmetric waves are excited in the pipe and then the reflected flexural and axisymmetric modes are analysed to estimate the location and size of a defect. Hayashi et al. \cite{Hayashi-2005} developed a method to analyse reflected axisymmetric and flexural modes. They reconstruct an image of the defect by a time-reversal technique. Therefore, analyzing and implementing flexural modes is a crucial element of active and synthetic focusing methods of guided waves in pipe.\\
It has been shown that the simultaneous excitation of a ring of transducers around a pipe can excite quasi-axisymmetric modes very effectively. As we mentioned before, in the ray-plate method, flexural modes are the same as axsisymmetric modes, but they move at an angle corresponding to their flexural order. Considering this concept of flexural mode propagation suggests an angular excitation of the pipe can excite a pure flexural mode. Therefore, a normal load helical excitation (fig.\ref{fig:loadsubfig2}) is proposed to excite a special flexural mode. The helix angle is determined by the ray-plate method. Different angles of the helix can excite different flexural orders. The helix angle $\alpha$ is determined by using eq. \eqref{PBC} as follows:
\begin{equation}\label{eq:Angles}
\alpha=sin^{-1}\left( \frac{mc^{(a)}_p}{2\pi f R}\right),
\end{equation}
where $ f $ is frequency of excitation, $ c^{(a)}_p $ is the axisymmetric phase velocity of the desired family, and $ m $ is the order of the flexural mode that is excited.\\
Flexural mode excitation in a  $ 3.5[in] $ schedule 40 steel pipe is considered. Where the external diameter of the pipe is $ D=3.5 [in] $ and its thickness is $ d=0.22[in]$. Because of a clear separation of the flexural modes at lower frequencies, the frequency of excitation is chosen to be as low as $ 50[kHz] $. Considering the longitudinal dispersion curves of the pipe shows that at this frequency two axisymmetric modes $L(0,1)$ and $L(0,2)$ exist where their phase velocities are $ c_p=1525[m/s] $ and $ c_p=5676[m/s] $, respectively. Substituting these data in eq.\eqref{eq:Angles} gives table \ref{table:angles} of the ray-plate method predictions for the angle of propagation/excitation for $ m=0,1,2,3,4,5 $.
\begin{table}
\caption{
$L(m,1)$ and $L(m,2)$ propagation directions predicted by the ray-plate method}
\centering
\begin{tabular}{|c|c|c|c|c|c|c|}
\hline f=50[kHz] & m=0  & m=1  & m=2 & m=3  & m=4 & m=5 \\
\hline  $ \alpha_{L(m,1)}  [deg]$   & 0	& 6.27 & 12.61 & 19.12 & 26.0& 33.1 \\
\hline  $ \alpha_{L(m,2)} [deg]$  & 0	& 24 &  54.38 & ---  & --- & --- \\
\hline
\end{tabular}
\label{table:angles}
\end{table}
Furthermore, phase velocities are predicted as table \ref{table:phasevelocity}  by the ray-plate method.
\begin{table}
\caption{$L(m,1)$ and $L(m,2)$ phase velocities predicted by the ray-plate method}
\centering
\begin{tabular}{|c|c|c|c|c|c|c|}
\hline f=50[kHz] & m=0  & m=1  & m=2 & m=3  & m=4 & m=5 \\
\hline  $ c^p_{L(m,1)}  [m/s]$  	& 1525	& 1534 & 1563 & 1614 & 1695 & 1820 \\
\hline  $ c^p_{L(m,2)}  [m/s]$  	& 5676	& 6212 &  9746 & ---  & --- & --- \\
\hline
\end{tabular}
\label{table:phasevelocity}
\end{table}
Finite element analysis is utilized for considering the flexural mode excitation by the proposed method. A helically normal edge load is applied on a  $ 3.5[in] $ schedule 40 steel pipe (see fig.\ref{fig:loadsubfig2}). The normal to the helix makes an angle of $ 20[deg] $ with respect to the z-axis. Table \ref{table:angles} shows that two flexural modes $ L(3,1) $ and $ L(1,2) $ have propagation angle $ \alpha=19.12[deg] $ and $ \alpha=24[deg] $, respectively. These angles are close to the excitation angle $ 20[deg] $. Therefore we expect to observe both  $ L(3,1) $ and $ L(1,2) $, but higher intensity at $ L(3,1) $ with phase velocity $ 1624 [m/s] $. A $ 50[kHz] $ Gaussian modulated signal is used as an excitation signal.

\begin{figure}
\centering
\mbox{
\subfigure[]{
   \includegraphics[height=.1\textheight,width=.8\textwidth] {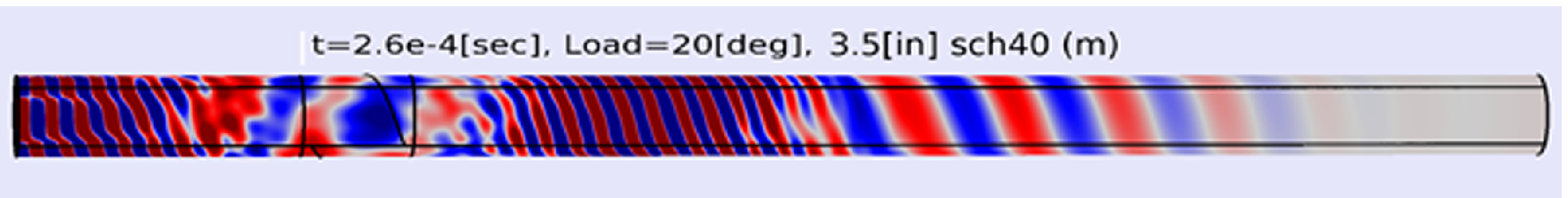}
   \label{fig:excitedsubfig1}
 }
 }
\mbox{
 \subfigure[]{
   \includegraphics[height=.35\textheight,width=.9\textwidth] {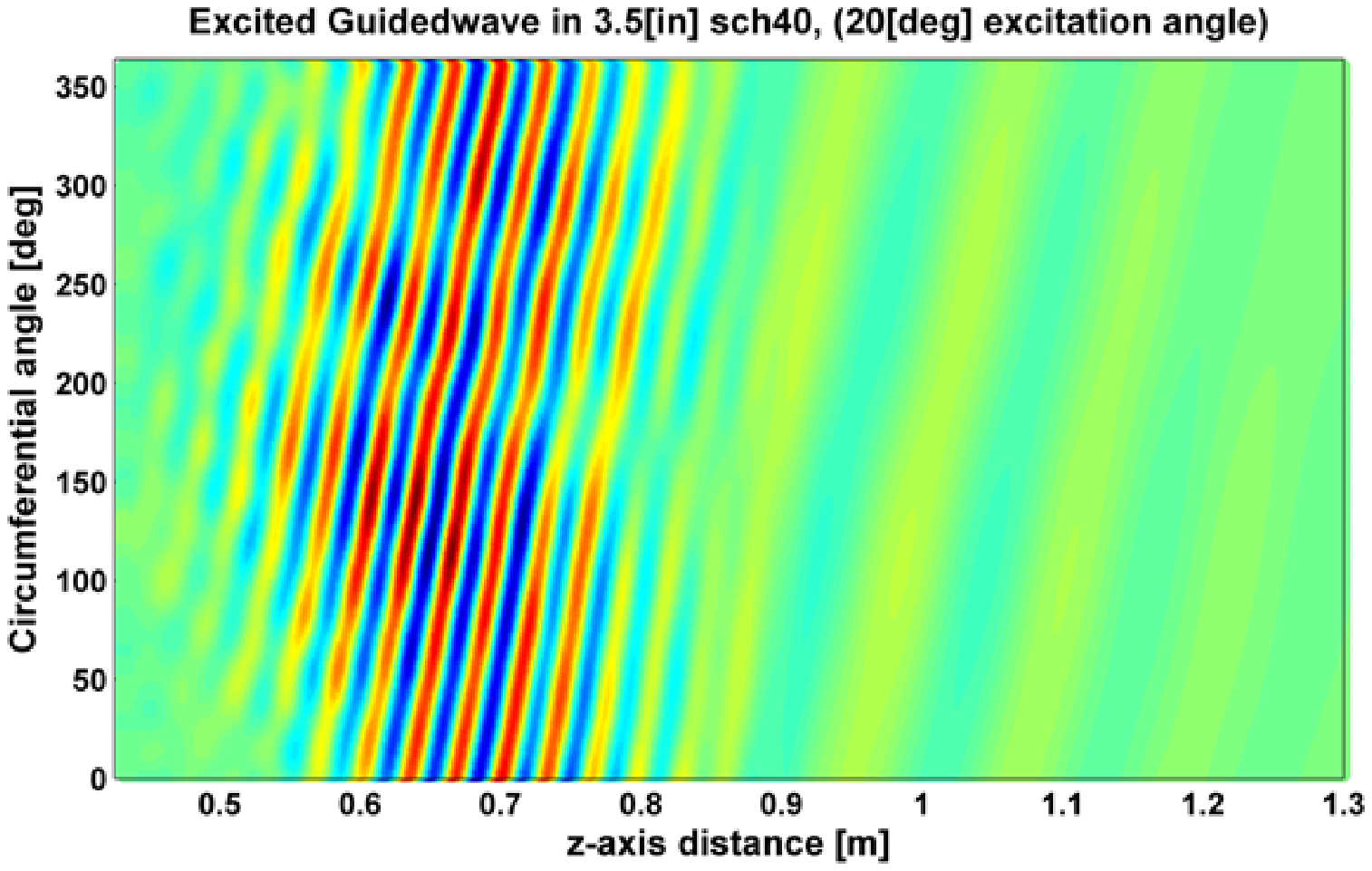}
   \label{fig:unwarpedsubfig2}
   }
 }
\label{fig:Wave}
\caption{Figures show the excited wave in a 3.5[in] schedule 40 steel pipe by $ 20[deg] $ helical excitation. a) shows directional propagation of the wave in the pipe, b) shows unwrapped view of the propagating wave. }
\end{figure}

\begin{figure}
\centering
\mbox{
\subfigure[]{
   \includegraphics[height=.4\textheight,width=.9\textwidth] {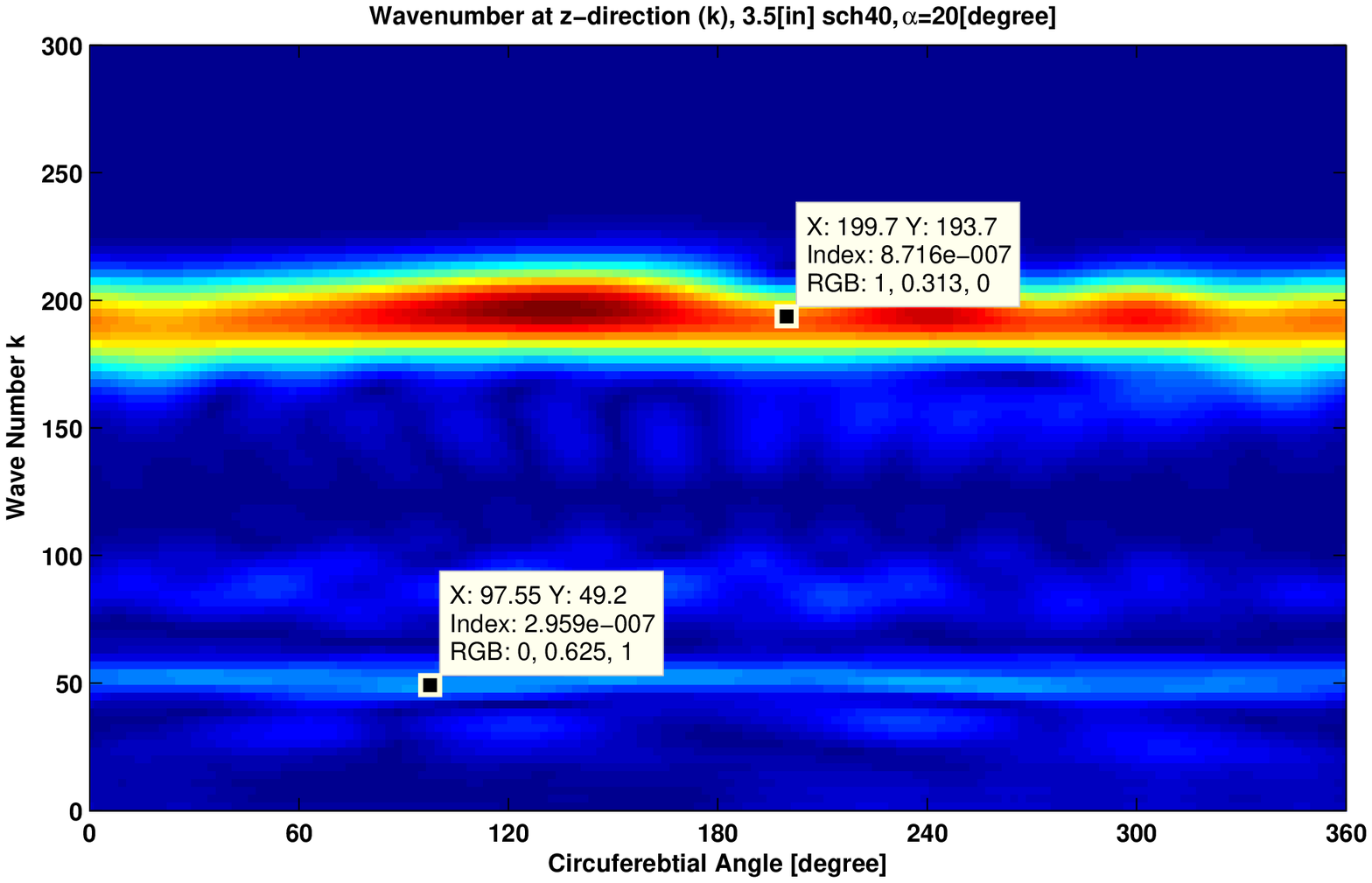}
   \label{fig:FFTsubfig1}
 }
 }
 \mbox{
 \subfigure[]{
   \includegraphics[height=.4\textheight,width=.9\textwidth] {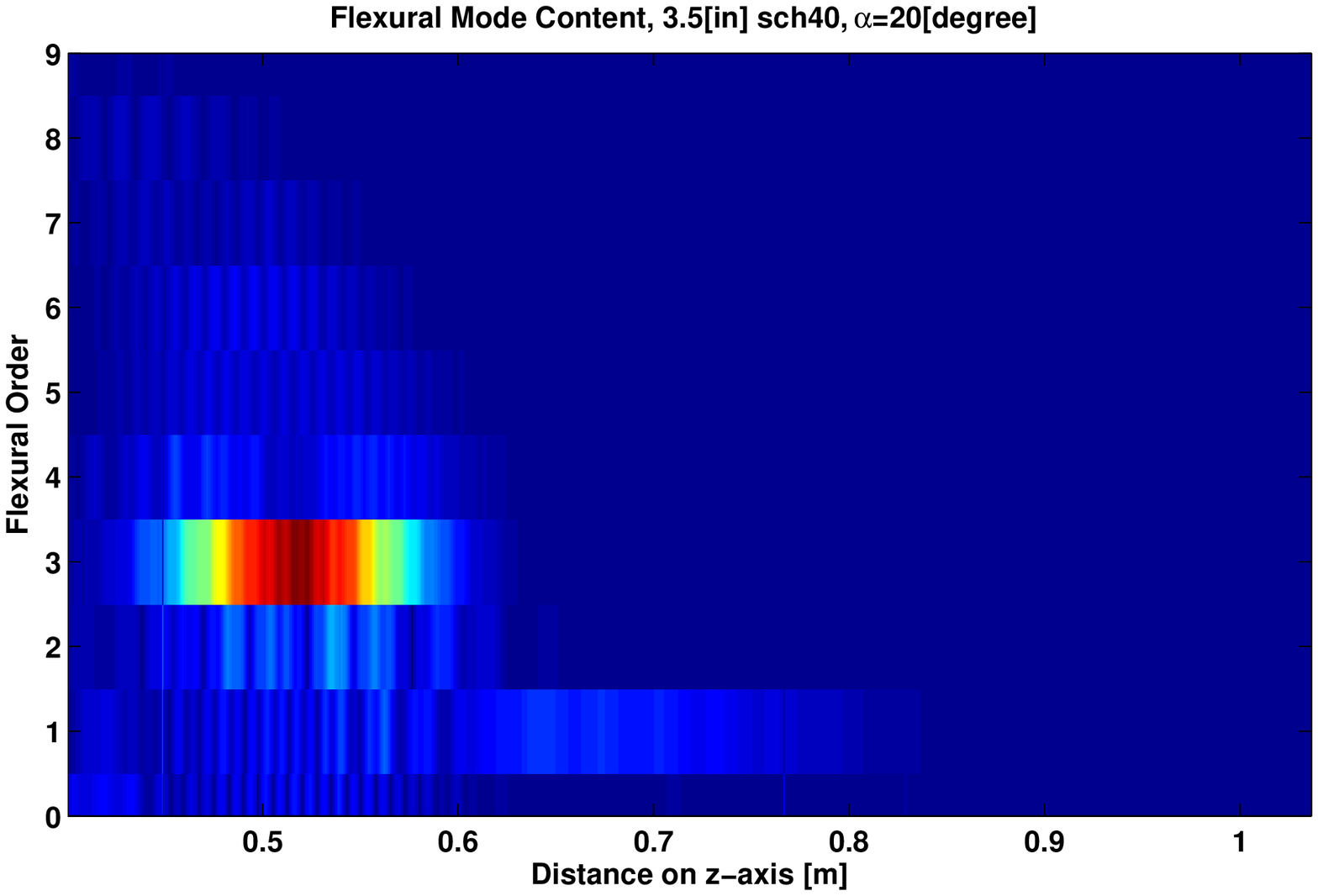}
   \label{fig:FFTsubfig2}
   }
}
\label{fig:FFT1}
\caption{ Figure (a) shows the wavenumber $ k $ of excited modes. There are two distinct wavenumbers $ k=193.7[m^{-1}] $ and $ k=49.2[m^{-1}] $ corresponding to $ c_p=1621[m/s] $ and $ c_p=6385[m/s] $, respectively. According to table\eqref{table:phasevelocity}, these phase velocities  correspond to $ L(1,3) $ and $ L(2,1) $. }
\end{figure}

\begin{figure}
\centering
\mbox{
\subfigure[]{
   \includegraphics[height=.4\textheight,width=.9\textwidth] {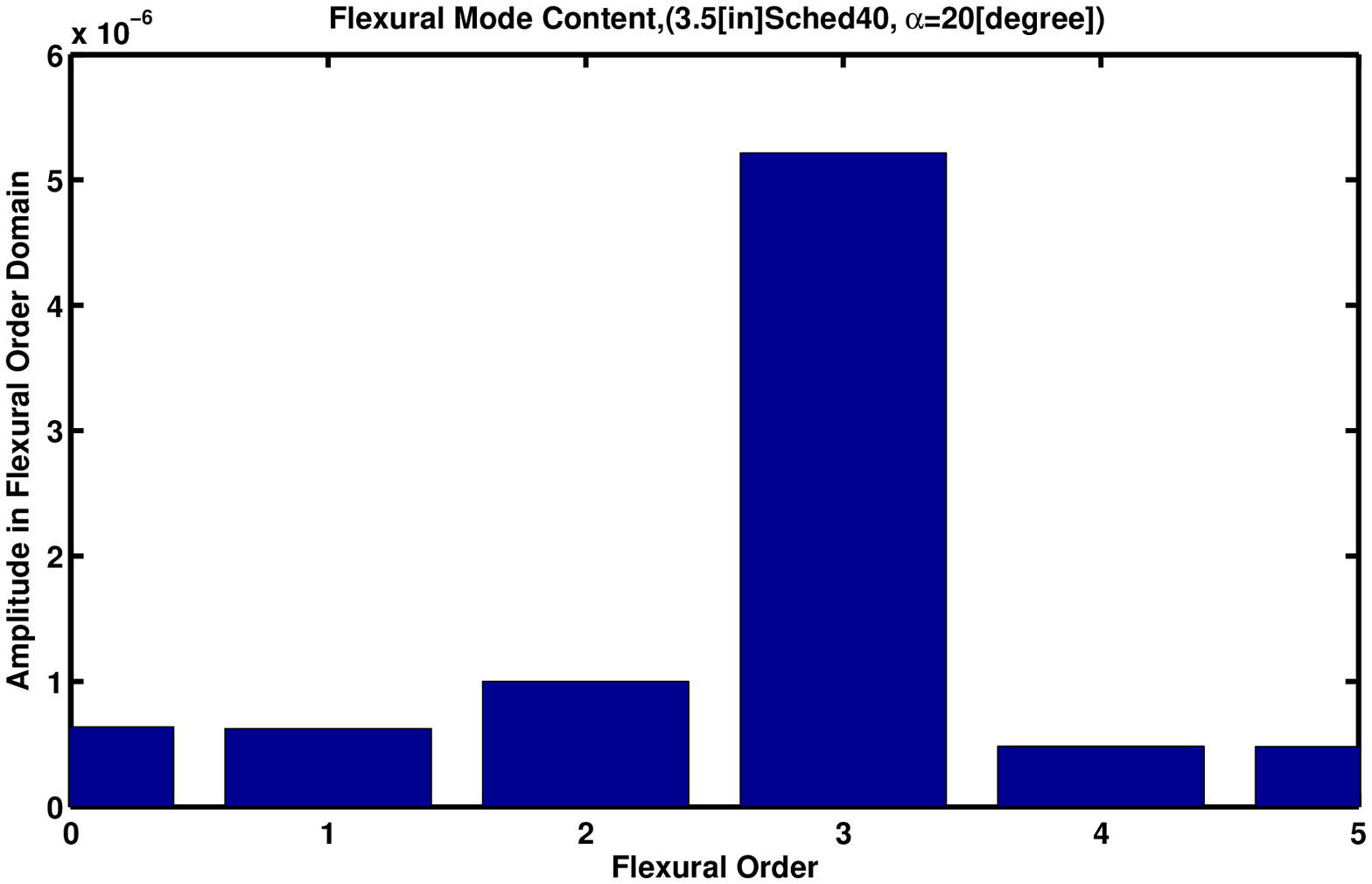}
   \label{fig:Flexordersubfig1}
 }
 }
 \mbox{
 \subfigure[]{
   \includegraphics[height=.4\textheight,width=.9\textwidth] {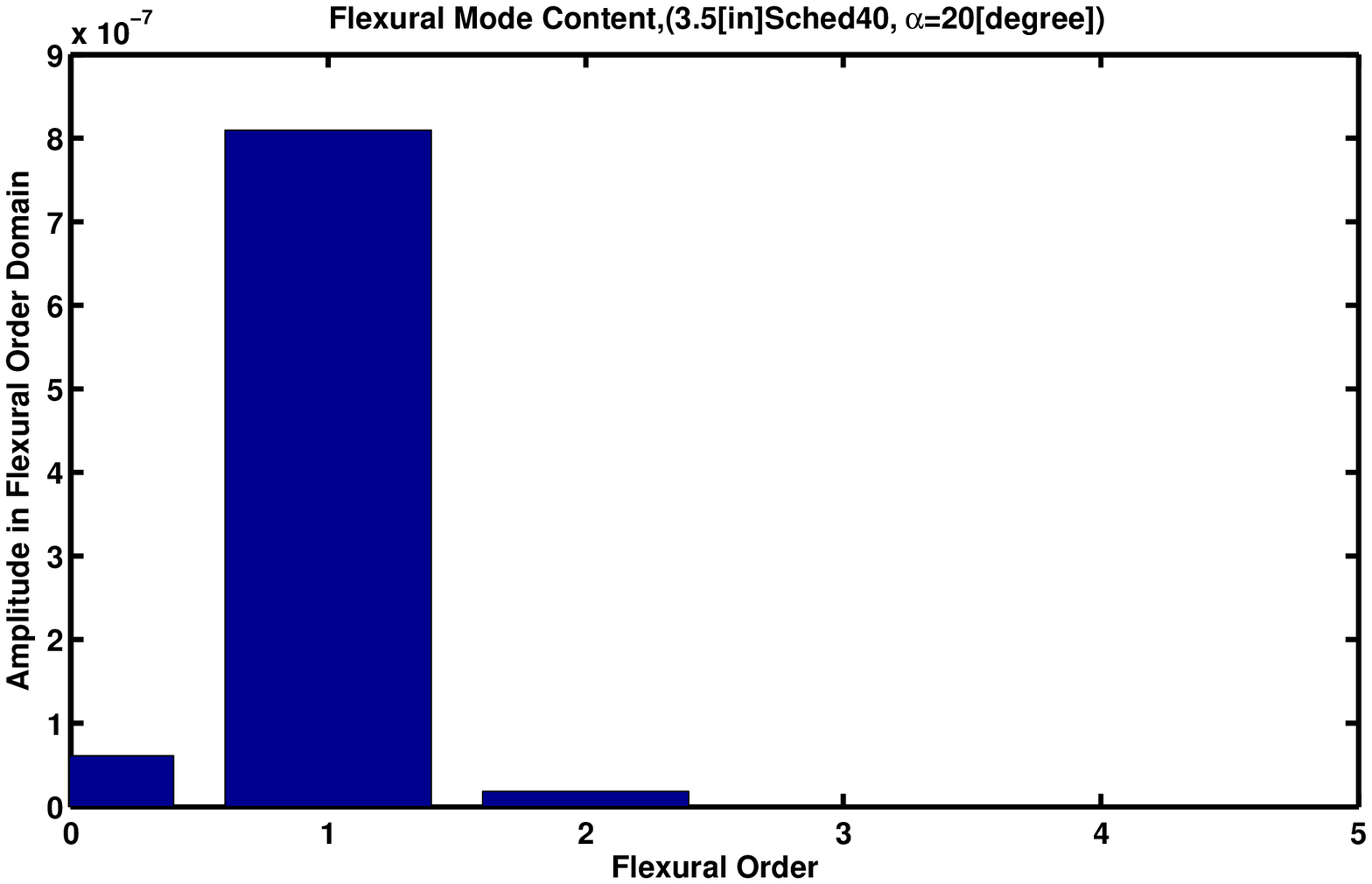}
   \label{fig:Flexordersubfig2}
   }
 } 
\label{fig:FFT3}
\caption{ Figure shows the order of excited flexural modes at $ z=0.65[m] $ and $ z=1[m] $ at $ t=0.26[msec] $. a) shows that at $ z=0.65[m] $ flexural mode has order three. b) shows that at $ z=1[m] $ a flexural mode of order one has been excited. }
\end{figure}

Figures \ref{fig:excitedsubfig1} and \ref{fig:unwarpedsubfig2} show excited waves in the pipe at $ t=0.26[m sec] $ following the initiation of the excitation. The propagation direction of the wave clearly makes an angle with the z-axis. In addition, the wave is divided into two parts because of a difference in the group velocities. Figure \ref{fig:unwarpedsubfig2} shows the unwrapped version of the excited waves in the pipe.
Phase velocities and flexural orders of these waves can be derived by taking the Fourier transform of the wave displacement field components on the z-axis and the pipe circumference. The z-component of the displacement field on the external surface of the pipe is recorded. Taking the Fourier transform of these data on the z-direction gives us the wavenumber $ k $. Then, using $ c_p=\omega/k $, the phase velocity of the wave  around the pipe can be calculated. Figure \ref{fig:FFTsubfig1} shows that there are two distinct wave numbers  $ k=193.7[m^{-1}] $ and $ k=49.2[m^{-1}] $ corresponding to $ c_p=1621[m/s] $ and $ c_p=6385[m/s] $, respectively. In the next step, the Fourier transform of the data is taken in the circumferential direction to extract the flexural orders. Figure \ref{fig:FFTsubfig2} shows the order of the excited flexural modes at all axial distances. Figure 8 shows the order of the excited flexural modes at $ z=0.65[m] $ and $ z=1[m] $. Figure \ref{fig:Flexordersubfig1} shows that at $ z=0.65[m] $ the flexural mode is of order three. Figure \ref{fig:Flexordersubfig2} shows that at $ z=1[m] $ a flexural mode of order one has been excited. Considering the derived phase velocities for the excited modes in fig. \ref{fig:FFTsubfig1}, the presented flexural order in figs. \ref{fig:FFTsubfig2}, \ref{fig:Flexordersubfig1}, \ref{fig:Flexordersubfig2} along with tables \ref{table:angles} and \ref{table:phasevelocity}  suggest that the flexural modes $ L(3,1) $ and $ L(1,2) $ have been excited effectively. 
In conclusion, the FEM proves that the proposed method for flexural mode excitation works quite well. This implies that the interpretation of flexural modes by the ray-plate method is correct. In addition, the predicted angles for flexural mode propagation by the ray-plate method are approximately correct. Therefore, the results in this section demonstrate the correctness and usefulness of the ray-plate method once more.
\section{Summary}
\label{Con}
A new method was developed for the consideration of wave propagation on complex waveguide surfaces. The method was derived from a phenomenological point of view and is computationally simple enough to apply to complex but practical situations. The method was extended to thick-walled waveguides where the thickness of the structure partially determines the dispersion and propagation properties of the guided waves. Pipe dispersion curves were used to prove the accuracy and the applicability of the method. For the first time, approximate analytical relations were derived for torsional and longitudinal dispersion curves. A physical interpretation of flexural modes in pipes was presented. Based on this new understanding of flexural modes, a helical excitation method was proposed for flexural modes. Finite element analysis was utilized to show that helical excitation works appropriately. FEM showed that as the ray-plate method predicts, a $ 20[deg] $ helical excitation can effectively excite $ L(3,1) $ and $ L(1,2) $.  In addition, limitations of the ray-plate method were considered for structures with  the small radius of curvature. It was shown that the predictions of the method are accurate if the ratio of the thickness to the mean radius is less than $ 0.062 $.\\
Continue this work, the proposed flexural mode excitation method is under study from several aspects. First, alternative methods to helical excitation are considered like using time delayed phased array. Also, the ability of ray-plate method to excite pure flexural modes at a wide range of frequencies and phase velocities is considered.
\section{References}
\label{Con}

\end{document}